\documentclass{aa}
\usepackage{astron}
\usepackage{graphics}
\usepackage{psfig}
\usepackage{times}

\def\etal{et al.}
\def\hii{H{\sc ii}}
\def\cbeta{$c_{\rm H\beta}$}

\def\rvincin{$r_{\rm 25}$}
\def\rtrois{$R_{\rm 3}$}
\def\rdeuxtrois{$R_{\rm 23}$}

\def\logrtrois{log$(R_{\rm 3})$}

\def\micron{$\mu$m}

\def\kmsMpc{km s$^{-1}$ Mpc$^{-1}$}

\def\ergs{erg s$^{-1}$}

\def\magarc{mag arcsec$^{-2}$}
\def\halpha{H$\alpha$}
\def\hbeta{H$\beta$}
\def\oii{[O\,{\sc ii}]$\lambda$3727}
\def\oiii{[O\,{\sc iii}]}
\def\oiiia{[O\,{\sc iii}]$\lambda$4959}
\def\oiiib{[O\,{\sc iii}]$\lambda$5007}
\def\Nii{[N\,{\sc ii}]}

\def\niib{[N\,{\sc ii}]$\lambda$6584}

\def\oiiishb{[O\,{\sc iii}]/H$\beta$}

\begin{document}
\thesaurus{03(11.01.1; 11.19.2; 11.19.3)}
\title{Starbursts in barred spiral galaxies}

\subtitle{IV. On young bars and the formation of abundance gradients
\thanks{Based on observations obtained at the 193cm telescope of
Observatoire de Haute--Provence, operated by INSU (CNRS)}
}

\author{S. Consid\`ere \inst{1}
       \and R. Coziol \inst{1}
       \and T. Contini \inst{2}
       \and E. Davoust \inst{3}
}

\offprints{considere@obs--besancon.fr}

\institute{Observatoire de Besan\c{c}on, UPRES--A 6091, B.P. 1615,
            F--25010 Besan\c{c}on Cedex, France
            \and
            European Southern Observatory, Karl--Schwarzschild--Strasse 2,
             D--85748 Garching bei M\"unchen, Germany
            \and
            Observatoire Midi--Pyr\'en\'ees, UMR 5572, 14 Avenue E. Belin,
            F--31400 Toulouse, France
          }

\date{Received 20/09/1999; accepted 14/01/2000}

\authorrunning{Consid\`ere et al.}
\titlerunning{IV. On young bars and the formation of abundance gradients}
\maketitle

\begin{abstract}

The oxygen (O/H) and N/O abundance ratios along the bar of 16
barred spiral starburst galaxies are determined using long--slit
spectroscopy.  The abundance gradients and the spatial
distribution of the ionized gas along the bar are used to
understand the role of bars in starburst galaxies.

The oxygen abundance gradients are steeper than in normal barred galaxies,
with a mean of --0.15 dex/kpc, while the intersects are low.  This excludes
the possibility that starburst galaxies in this sample are chemically
evolved galaxies rejuvenated by the effect of a bar.
The nitrogen-to-oxygen abundance gradients are
flatter than the oxygen ones with a mean of --0.05 dex/kpc.  But N/O
intersects are high, which rules out the possibility that a large quantity of
gas was recently funneled by a bar toward the center of a young galaxy.
There is
no correlation between the abundance gradients and the bar properties, which
suggests that bars did not influence the chemical evolution of these galaxies.
Therefore, bars cannot be at the origin of the bursts in the nuclei of
our sample galaxies.

The oxygen and N/O abundance gradients are generally stronger in the
bar than in the disk
and are linked together by a linear relation consistent with a $primary +
secondary$ origin for the production of nitrogen.  This can be fully explained
in terms of star formation history in these galaxies. The
gradients build up from the inside out, becoming stronger as the oxygen and
N/O abundances increase in the bulge while staying low in the disk.
This behavior is consistent with a simple Schmidt law relating
the density of star formation to that of gas.

In many of the sample galaxies, star formation occurs at one or both ends
of the bar.
The low level of chemical enrichment in these regions suggests that they
recently experienced bar-triggered star formation: this is the only
visible effect of bars.
Our analysis shows that bars
probably appeared very recently (a few $10^7$ years) in the starburst
galaxies, which are relatively ``young'' galaxies still in the process of
formation.

\end{abstract}



\section{Introduction}

Massive galaxies play a fundamental role in the chemical evolution
of the Universe, because they are the only systems which meet the physical
conditions necessary for the build--up of heavy elements.
Today, by using simple techniques and modest--size telescopes, it is
possible to understand how the production of metals occurs in
nearby massive galaxies and to relate this production to their formation and
evolution processes.

While still in its early stages, the study of chemical abundances in
external galaxies has already produced important results. One of the most
interesting is the discovery that galaxies in the Universe seem to
follow a mass--metallicity relation \cite{ZARITSKYetal94}. Although we still
have to understand the physical causes behind this behavior, there is no doubt
that this phenomenon is a key parameter for understanding galaxy evolution.
With this prospect in mind, we recently verified that the massive starburst
nucleus galaxies (SBNGs) follow the same mass--metallicity relation as normal
ones \cite{COZIOLetal97a}. Our observations not only confirm the universality
of the mass--metallicity relation for a wide range of galaxies, but also
suggest that its origin may be linked to the different paths followed by early
and late--type spiral galaxies to form their bulge and
disk \cite{COZIOLetal98}.

Another important phenomenon, unveiled by chemical
abundance studies, is that the
metallicity in the ionized interstellar medium of spiral galaxies decreases
outward (e.g. Vila--Costas \& Edmunds, 1992, and references therein). Various
hypotheses have been proposed to explain this gradient \cite{P89,GK92}, but
their verification is difficult. The reason is that many parameters and
processes involved in the formation of an abundance gradient are still not well
understood, such as the initial mass function, the yield, the mechanisms and
time scales for the formation of halos, bulges and disks, as well as the
possible infall or outfall of matter. Different numerical simulations have
shown, however, that one important ingredient for producing abundance
gradients is a non linear dependence of the star formation rate on gas density
\cite{WS89,GK92,MOLLAetal96}. It has also been suggested that once an
abundance gradient is established, the presence of gas flows can either
amplify or reduce it \cite{EDMUNDS90,GK92}.

Attempts to relate abundance gradients to other characteristics of galaxies,
such as morphological type, luminosity or circular velocity, have also been
performed \cite{V-CE92,OK93,ZARITSKYetal94}, but no clear correlation has
emerged. One interesting trend is however observed: barred spiral galaxies
seem to
have shallower abundance gradients than non--barred ones
\cite{V-CE92,EdR93,ZARITSKYetal94,MR94}. From a theoretical point of view,
bars are expected to funnel gas from the outer parts to the nuclei of galaxies
\cite{N88}. The flow along the bar can induce strong mixing
\cite{FRIEDLIetal94} and reduce metallicity gradients. But star formation
induced by the bar can also be strong enough to counteract dilution effects
and increase the gradient in the inner parts of galaxies. Steep abundance
gradients have indeed been reported in the inner parts of some barred
galaxies, such as NGC 3359 \cite{MR95} and NGC 1365 \cite{RW98}.

In the present paper, we analyse the abundance gradients of oxygen
(O/H) and nitrogen to oxygen ratio (N/O) in 16 barred starburst
galaxies. Our goal is to establish if starburst galaxies follow
the trends observed in normal galaxies. We also want to determine
if there is a relation between the bar and the burst of star
formation. We have recently proposed that SBNGs may be ``young''
galaxies still in their formation process (Coziol \etal\
1997a,1998). This possibility raises several new questions. Is the
initial distribution of abundances in starburst galaxies flat, or
have they already had enough time to establish steep gradients? Is
it possible to determine how the gradient builds up in a young
galaxy and establish whether the presence of the bar is important?
Could bars be younger in starburst than in normal galaxies?

Our analysis innovates by studying the abundance gradients of two
key elements, oxygen and nitrogen, which are assumed
\cite{COZIOLetal99a} to be produced by different types of stars
and, thus, to be released into the interstellar medium on
different time scales. This gives us a deeper insight into the
processes of chemical evolution, because we can test the influence
of star formation and of structures, like a bar, over a relatively
long period of time (a few Gyrs). This advantage should in
particular enable us to determine whether bars are young and if
they play a major role in triggering or feeding the nuclear
starbursts.

\section{The sample of barred starburst galaxies}

\begin{table}
\caption[]{Catalogue elements of the sample galaxies
and excerpts from our observing logbook}
\small
\begin{flushleft}
\begin{tabular}{lrlrrrrrr}
\noalign{\smallskip} \hline \hline \noalign{\smallskip}
\multicolumn{2}{c}{Galaxy}&Type &$D$& \rvincin&Exp.& P.A.\\
&&&[Mpc]&[\arcsec]&[min]&[deg]\\ \noalign{\smallskip} \hline
\noalign{\smallskip} Mrk&   2 &SB0/a&76&23.3&   45&  66\\ Mrk&  12
&SABc &56&36.1&   40&  32\\ Mrk&  13 &SBb  &22&34.4&   40& 128\\
Mrk& 306 &SBbc &77&36.1&   90&   7\\ Mrk& 307 &SBc  &76&34.4&
90&  22\\ Mrk& 326 &SABbc&49&52.1&   45& 111\\ Mrk& 332 &SBc
&34&44.3&   90& 100\\ Mrk& 373 &SBc  &81&21.7&  90&  20\\ Mrk& 545
&SBa  &63&64.1&   90& 159\\ Mrk& 602 &SBbc &38&40.5&   90&  33\\
Mrk& 710 &SBab &20&67.1&   30& 33\\ Mrk& 712 &SBbc &62&33.7&   45&
8\\ Mrk& 799 &SBb  &43&68.7& 3$\times$25& 144\\ Mrk& 898 &SBbc
&70&32.9&   45& 147\\
Mrk&1076 &SBbc &96&24.4&   90& 148\\ Mrk&1088 &SB0/a&61&55.9&
90&  25\\ NGC&6764 &SBbc &36&73.6&16,25& 62,74\\
\noalign{\smallskip} \hline \hline
\end{tabular}
\end{flushleft} \label{TPARAM}
\end{table}

The 16 galaxies studied in this paper are a sub--sample of 144 Markarian
barred spiral galaxies studied spectroscopically by Contini \cite*{C96} and
Contini \etal\ \cite*{CONTINIetal98} (Paper III). This sub--sample
is a selection of galaxies which are experiencing very intense bursts of
star formation, as estimated by their high \halpha\ luminosity (L(\halpha) $>
10^{40}$ \ergs) or large \hbeta\ equivalent width (EW(\hbeta) $>$ 30 \AA).
Three Wolf--Rayet galaxies are included in our sample: Mrk 710, 712 and 799.
Another Wolf--Rayet galaxy satisfying our selection criteria, NGC 6764,
\cite{CONTINIetal97}, was later added to the list; but, as we will show later,
it turns out to be a LINER and cannot be included in our analysis.

Some catalog elements for the sample galaxies and excerpts from our
observing logbook are presented in Table~\ref{TPARAM}. The morphologies were
taken in NED (http://nedwww.ipac.caltech.edu), or in LEDA
(http://www--obs.univ--lyon1.fr) for Mrk 13 and 306. Table~\ref{TPARAM}
also lists
the distance $D$, estimated from the recession velocity (assuming H$_o = 75$
\kmsMpc), and the radius \rvincin\ at 25 \magarc. Both values were found in
LEDA. The last two columns indicate the exposure times and position angles of
the slit which were used during our spectroscopic observations.

\section{Spectroscopic observations and reductions}

Most of the spectroscopic observations were obtained in October 1995 and June
1996 at the 1.93 meter telescope of Observatoire de Haute--Provence. The CCD
was a thinned 512$\times$512 Tektronix (pixel size 27\micron). We used the
Carelec spectrograph \cite{LEMAITREetal90} with a spectral dispersion of 260
\AA/mm, which covers the spectral range 3600 to 7200\AA, at a resolution of
$\sim 7$\AA. The slit was aligned along the bar. During the first run, we took
90--minute spectra of most galaxies, with a slit width of 3.0\arcsec. For flux
calibration, we observed the standard stars G191B2B, Hilt 600, HD 217086 and
BD +28$^{\circ}$ 4211 \cite{MASSEYetal88}. In the second run, we took three
25--minute spectra of Mrk 799 and two of NGC 6764, with a slit width of
2.8\arcsec. One standard star, HD 192281, was used for flux calibration
\cite{MASSEYetal88}. The galaxies Mrk 710 and 712 were observed during
previous runs at the same telescope, using similar instrumental settings
\cite{CONTINIetal98}. The average seeing during the nights was of the order of
1.2\arcsec.

All spectra were reduced with MIDAS, applying standard procedures: offset
and flatfield corrections, sky subtraction, wavelength and flux calibration,
airmass and galactic extinction corrections and elimination of cosmic impacts.
In order to study abundance variations along the bar, series of
one--dimensional spectra were extracted from the two--dimensional spectra of
each galaxy by averaging three contiguous rows along the slit. Observed
emission--line fluxes and equivalent widths were then measured in each
elementary spectrum. For the rest of the reduction we followed the procedure
outlined in Contini \etal\ \cite*{CONTINIetal95}\ (Paper I): the Balmer
(\halpha\ and \hbeta) absorption contamination due to underlying stellar
populations was corrected by
adding 2\AA\ to the equivalent width of the Balmer emission lines;
the principal Balmer
decrement (\halpha/\hbeta) was used to estimate the reddening coefficient
\cbeta\ and to calculate absolute dereddened fluxes.

Our analysis may differ from other studies found in the literature, because we
measured the spatial variations of physical parameters only along one direction
(along the bar), and not across the whole two--dimensional disk.  This
difference should be kept in mind when comparing our results with those of other
studies.

\section{Properties of emission--line regions}

\begin{figure}
\resizebox{\hsize}{!}{\includegraphics{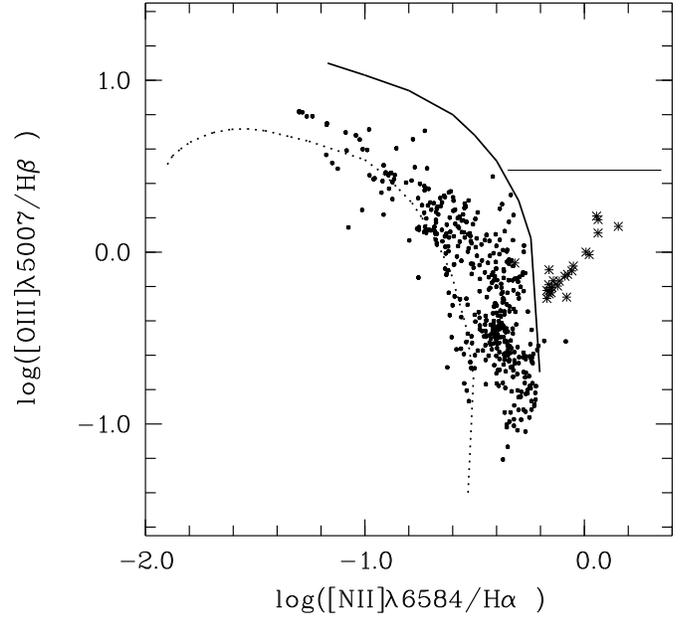}}
\caption{
Diagnostic diagram for the emission--line regions of the 17 sample galaxies.
The curved solid line separates \hii\ region--like objects ({\it left})
from active galaxies ({\it right}). The horizontal solid line separates
Seyfert 2 nuclei ({\it top}) and LINERs ({\it bottom}). The dotted curve
denotes the theoretical model of disk \hii\ region of McCall \etal\ (1985).
Most of the emission--line regions are of \hii\ type, except  those located in
the central part of NGC 6764 which are of LINER type ({\it asterisks})
}
  \label{diag}
\end{figure}

A standard diagnostic diagram \cite{BPT81,VO87} was used to determine the
dominant source of excitation of the emission--line regions in our sample. Most
of these regions fall in the domain of \hii\ region--like spectra
(Fig.~\ref{diag}), referred to hereafter as \hii\ regions. The only exceptions
are two regions in Mrk 545, which have weak signal--to--noise ratios, and the
inner region of NGC 6764. There has been some confusion in the literature
about the nature of the activity in this last galaxy (see Contini \etal\ 1996
for the various classifications).  Our new spectra clearly show that the
center of NGC 6764, as well as the circum--nuclear regions, are the seat of
LINER activity. Because LINER is a non--thermal phenomenon (implying either an
AGN or shocks), the metallicity in these regions cannot be determined by the
methods used in this paper. Consequently, NGC 6764 (and the two regions in Mrk
545) have been excluded from our abundance gradient analysis.

Our sample of \hii\ regions spans a large range of excitation levels, from
high--excitation spectra (log (\oiiishb) $\geq$ 0.4), characteristic of \hii\
galaxies, to low--excitation spectra (log (\oiiishb) $<$~0.4) typical of
SBNGs \cite{COZIOL96}. The low--excitation \hii\
regions have \niib/\halpha\ ratios which are relatively high.  This is a
common property of SBNGs \cite{COZIOLetal97b,CONTINIetal98}, which possess a
slight overabundance of nitrogen as compared to \hii\ regions with comparable
metallicity \cite{COZIOLetal99a}.

\section{Determination of oxygen and nitrogen abundances}
\label{METALLI}

\subsection{Determination of oxygen abundances: general
method}

The oxygen abundances are determined using empirical relations between
log(O/H) and the emission--line ratios \rtrois\ and \rdeuxtrois, which are
defined as:

\begin{eqnarray}
\mbox{\rtrois} = (\mbox{\oiiia} + \mbox{\oiiib})/\mbox{\hbeta}
\end{eqnarray}
\begin{eqnarray}
\mbox{\rdeuxtrois} = (\mbox{\oii} + \mbox{\oiiia} + \mbox{\oiiib})/\mbox{\hbeta}
\end{eqnarray}

For our analysis we use the empirical relations first established by Edmunds
\& Pagel \cite*{ED84} and quantified by Vacca \& Conti \cite*{VC92}.  In
general, the above two relations yield similar results. In some cases,
however, the relation based on \rtrois\ gives higher oxygen abundances than the
one based on \rdeuxtrois, because the extinction correction affects
differently the two relations; a high extinction correction tends to decrease
the ratio \rtrois, while increasing the ratio \rdeuxtrois. The consequence of
this effect is that when dust extinction is high, one overestimates the
oxygen abundance using \rtrois\ and underestimates it using \rdeuxtrois. For
this reason, we will use both relations and adopt the mean value. Exceptions
are Mrk 710 and 712, where \oii\ was not observed and only \rtrois\ was used.

\subsection{Determination of extinction in the center of galaxies}
\label{BALMABS}

In the center of half of the sample galaxies (Mrk 12, 307, 326, 332,
545, 799, 898 and 1088), Balmer absorption features due to an underlying
intermediate--age stellar population severely affect the measurement of the
Balmer emission lines. This prevents us from determining the extinction
coefficient \cbeta\, which is needed for calculating the abundances.  The
standard correction of 2\AA\ on the equivalent width is not adequate for
these cases (because we still cannot measure \hbeta\ correctly) and it is thus
necessary to find a more suitable method to solve this problem.

It has been known for some time that there is an empirical relation between
\logrtrois\ and log(\Nii/\oiii) \cite{ALLOINetal79,ED84}. The advantage of
using the \Nii\ and \oiii\ emission lines for determining the oxygen
abundance is that they are not affected by underlying stellar populations.
Such a method is ideal for analyzing the central regions of galaxies. However, we still have
to correct these lines for dust extinction. As we will now show, this can be
done empirically.

According to the definition of the extinction coefficient, the observed and
dereddened fluxes of the \niib\ and \oiiib\ lines (hereafter \Nii\ and \oiii)
are linked by the relation:

\begin{eqnarray}
\log{\mbox{\Nii/\oiii}}_{dred} = \log{\mbox{\Nii/\oiii}}_{obs}
-0.3 \times \mbox{\cbeta}
\end{eqnarray}

\begin{figure}
\resizebox{\hsize}{!}{\includegraphics{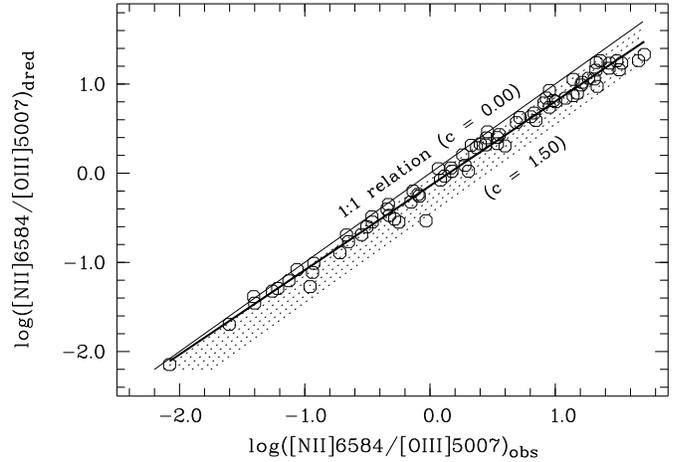}}
  \caption{
Dereddened {\it versus} observed flux ratios of \niib\ and \oiiib.
The thin line is a 1 to 1 relation. A set of dotted parallel lines represents
6 ranges of \cbeta\, varying from 0.25 to 1.50 (see text). The sample of
normal galaxies of McCall et al.  (1985) is used as an example ({\it open
circles}). The thick line  is the regression on the data
}
  \label{RAWDER}
\end{figure}

\noindent
If one plots log(\Nii/\oiii)$_{dred}$ {\it versus} log(\Nii/\oiii)$_{obs}$ for
a sample of spectra, statistically large enough to include evenly distributed
values of \cbeta\ between 0.0 and $\sim$ 1.5, one expects the data to cover
uniformly the region between the line of zero extinction and that of \cbeta\
$\sim$ 1.5. We notice instead that the data gather along an oblique line of
slope smaller than unity, which intersects the lines of constant \cbeta\ (see
Fig.~\ref {RAWDER}). This is observed in two samples of \hii\ regions in
normal galaxies \cite{MCCALLetal85,V-CE93}, as well as in our own sample of
starburst galaxies (excluding the spectra affected by Balmer absorption). This
correlation translates the fact that, on average, the larger the value of
log(\Nii/\oiii) the larger that of \cbeta. The reason behind this behavior is
that log(\Nii/\oiii) is correlated with oxygen abundance (see
Sect.\ref{NINDICATOR} and Fig.~\ref{r3_n2o3}), which in turn is correlated
with the amount of dust and consequently with internal extinction
\cite{Hetal98}.

This observed correlation provides us with an empirical way of estimating
\cbeta\ from the observed fluxes of \Nii\ and \oiii. Using the sample of
galaxies of McCall et al. (1985) and our own sample (excluding the spectra
with Balmer absorption), we find the following relation between
the observed and dereddened line ratios:

\begin{eqnarray}
\mbox{log(\Nii/\oiii)}_{dred} = 0.93 \mbox{log(\Nii/\oiii)}_{obs} - 0.16
\end{eqnarray}

\noindent
The correlation coefficient of this relation is 0.99 and the uncertainty on
log(\Nii/\oiii)$_{dred}$ is less than 0.1 dex.  This relation will be used in
Sect.~\ref{NINDICATOR} to calculate (\Nii/\oiii)$_{dred}$, and thus
\rtrois, in the center of galaxies affected by Balmer absorption.

Using the same samples, we find the following relation between \cbeta\ and
(\Nii/\oiii)$_{obs}$:

\begin{eqnarray}
\label{chbeta}
\mbox{\cbeta} = 0.45 \mbox{log(\Nii/\oiii)}_{obs} + 0.51
\end{eqnarray}

\noindent
The correlation coefficient is about 0.95 and the uncertainty is about 0.1 in
\cbeta. This relation will be used in Sect.~\ref{RDEUXTROIS} to deredden the
\oiiib, \oii\ and \halpha\ lines in the center of galaxies with strong
Balmer absorption, and hence to calculate \rdeuxtrois.

\subsection{Calculating \rtrois\ in the center of galaxies}
\label{NINDICATOR}

\begin{figure}
\resizebox{\hsize}{!}{\includegraphics{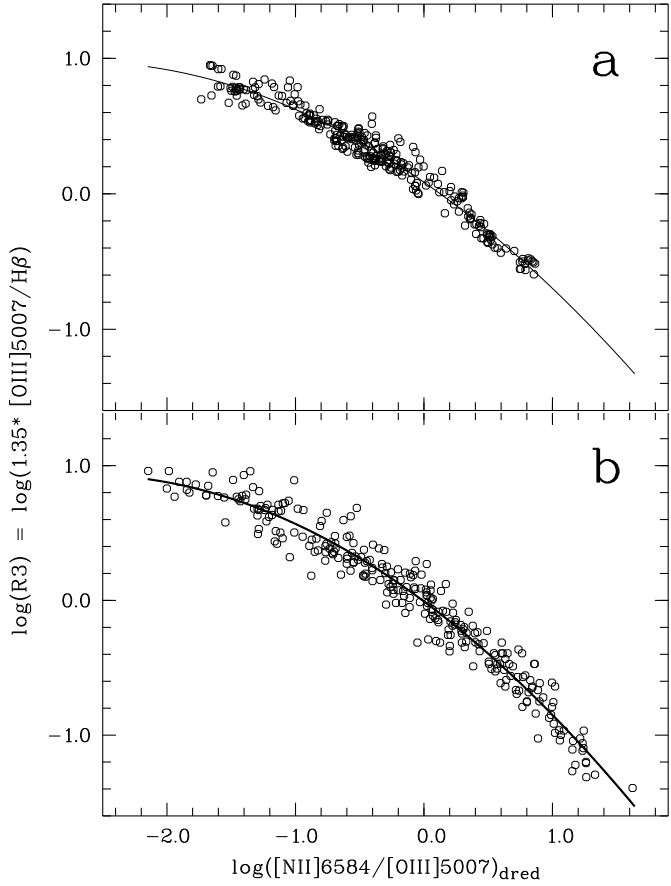}}
  \caption{
Relations between \logrtrois\ and dereddened log(\niib/\oiiib). {\bf a} for
the \hii\ regions in our sample of starburst galaxies that are not affected
by Balmer absorption.  {\bf b} for 278 \hii\ regions in normal
galaxies, from the samples of McCall et al. (1985) and Vila--Costas et al.
(1993). The two continuous lines in {\bf a} and {\bf b}
are adjusted polynomials
}
\label{r3_n2o3}
\end{figure}

Following an idea developed by various authors \cite{ALLOINetal79,ED84}, we
now search for an empirical relation between \rtrois\ and \Nii/\oiii\ in our
sample of 16 barred starburst galaxies. For this analysis we use only the
dereddened fluxes of \hii\ regions, about 300 of them, which are not affected
by Balmer absorption. The two quantities appear to be strongly correlated, as
shown on Fig.~\ref{r3_n2o3}a, and we fit a two--degree polynomial to the data,
which yields the following relation:

\begin{eqnarray}
\label{SBNG}
\mbox{\logrtrois}\ &=& -\ 0.123\  (\mbox{log(\Nii/\oiii)})^2 \nonumber \\
                   & & -\ 0.661\  \mbox{log(\Nii/\oiii)}\ +\ 0.086
\end{eqnarray}

\noindent
with a dispersion in \logrtrois\ equal to 0.07 dex.

To check the validity and the possible ``universality'' of the above
relation, we now apply the same procedure to two samples of \hii\ regions in
normal galaxies: those of McCall et al. (1985) and of Vila--Costas \& Edmunds
(1993).  As shown in Fig.~\ref{r3_n2o3}b, the behavior is the same. Fitting a
two--degree polynomial we obtain:

\begin{eqnarray}
\label{MRS}
\mbox{\logrtrois}\ &=& -\ 0.133\  (\mbox{log(\Nii/\oiii)})^2 \nonumber \\
                   & & -\ 0.709\ \mbox{log(\Nii/\oiii)}\ -\ 0.008
\end{eqnarray}

\noindent
The dispersion in \logrtrois\ is 0.12 dex, that is slightly higher
than for our sample.

It is important to note that the two fitted curves in Fig.~\ref{r3_n2o3}
differ by a horizontal shift of 0.14 dex. This shift is consistent with the
overabundance of nitrogen in SBNGs with respect to normal \hii\ regions
\cite{COZIOLetal99a}. Consequently, there is no unique relation between
\logrtrois\ and log(\Nii/\oiii), but specific relations for homogeneous
samples of galaxies, such as normal ones or SBNGs. For a given value of the
ratio \Nii/\oiii, a starburst galaxy will have a lower oxygen abundance than a
normal galaxy, or, for a given abundance, a starburst galaxy will have a
higher \Nii/\oiii\ ratio than a normal galaxy.

Figure~\ref{r3_n2o3}a shows that our data do not reach oxygen abundances
as high
as in normal galaxies.  This is precisely the domain where we cannot determine
\rtrois\ in the standard way, because, in starburst galaxies, such high values
for oxygen abundance are observed only in the nuclear regions, which are affected by
Balmer absorption features.  Fortunately, the parallel empirical relation for
normal \hii\ regions (Fig.~\ref{r3_n2o3}b) continues smoothly into the high
oxygen--abundance regime.  We therefore consider it legitimates to use
Eq.\ref{SBNG} to calculate \rtrois\ in this range of oxygen abundances.

\subsection{Calculating \rdeuxtrois\ and O/H in the center of galaxies}
\label{RDEUXTROIS}

Once we know \cbeta\ in the center of the sample galaxies
(using Eq.~\ref{chbeta}),
it is relatively easy to estimate \rdeuxtrois. First, we correct \oiiib,
\oii\ and \halpha\ for reddening using \cbeta\ and estimate the dereddened
\hbeta\ flux using the theoretical Balmer decrement: \hbeta$_{dred} =
\,$\halpha$_{dred} / 2.85$. Then, the dereddened ratio \rdeuxtrois\ is
computed assuming \oiii\ $= 1.35 \times$ \oiiib, and the oxygen abundance O/H
is deduced using the relation of Vacca \& Conti \cite*{VC92}.

Comparing the abundances obtained using \rtrois\ and \rdeuxtrois\ in the
center of galaxies, we find the same kind of sensitivity to dust
extinction as that observed in the outer parts of galaxies.
Consequently, we also adopt the average value of the oxygen abundances
estimated by \rtrois\ and \rdeuxtrois\ in the center of our sample galaxies.

\subsection{N/O abundance ratios}
\label{nitroabund}

\begin{table*}
\caption[]{Observed gradients and intersects in O/H and N/O}
{\scriptsize
\begin{flushleft}
\begin{tabular}{rcccccccc}
\noalign{\smallskip} \hline \hline \noalign{\smallskip} Galaxy
&\multicolumn{4}{c}{O/H}&\multicolumn{4}{c}{N/O} \\
            &\multicolumn{4}{c}{------------------------------------------------------------------------------------------}
            &\multicolumn{4}{c}{------------------------------------------------------------------------------------------} \\
            & Global   & Intersect& Bar     & Disk    & Global   & Intersect& Bar     & Disk    \\
 Mrk \#     & (dex/kpc)& [O/H]    &(dex/kpc)&(dex/kpc)& (dex/kpc)& N/O    &(dex/kpc)&(dex/kpc)\\
\noalign{\smallskip} \hline \noalign{\smallskip}\footnotesize
   2 & $ -0.034\pm0.022 $ & $ +0.31\pm0.03 $ & $ -0.034\pm0.022 $ &                   &
       $  0.000         $ & $ -0.63\pm0.07 $ & $  0.000         $ &                   \\
  12 & $ -0.093\pm0.007 $ & $ +0.16\pm0.08 $ & $ -0.071\pm0.008 $ & $ -0.081\pm0.010$ &
       $ -0.037\pm0.008 $ & $ -0.91\pm0.10 $ & $  0.000         $ & $ -0.041\pm0.012$ \\
  13 & $ -0.438\pm0.047 $ & $ -0.09\pm0.04 $ & $  0.000         $ & $ -0.510\pm0.082$ &
       $  0.000         $ & $ -1.14\pm0.04 $ & $  0.000         $ & $  0.000  $       \\
 306 & $ -0.041\pm0.007 $ & $ -0.18\pm0.07 $ & $ -0.078\pm0.004 $ & $  0.000  $       &
       $ -0.037\pm0.009 $ & $ -1.08\pm0.09 $ & $  0.000         $ & $  0.000  $       \\
 307 & $ -0.114\pm0.011 $ & $ +0.21\pm0.11 $ & $ -0.098\pm0.025 $ & $ -0.059\pm0.021$ &
       $ -0.040\pm0.008 $ & $ -0.76\pm0.09 $ & $  0.000         $ & $ -0.037\pm0.019$ \\
 326 & $ -0.269\pm0.027 $ & $ +0.49\pm0.10 $ & $ -0.305\pm0.047 $ & $  0.000$         &
       $ -0.154\pm0.023 $ & $ -0.56\pm0.11 $ & $ -0.124\pm0.048 $ & $  0.000$         \\
 332 & $ -0.169\pm0.017 $ & $ +0.39\pm0.12 $ & $ -0.175\pm0.056 $ & $ -0.242\pm0.028$ &
       $ -0.057\pm0.007 $ & $ -0.76\pm0.04 $ & $ -0.072\pm0.023 $ & $ -0.050\pm0.012$ \\
 373 & $ -0.030\pm0.012 $ & $ -0.12\pm0.05 $ & $ -0.045\pm0.018 $ &                   &
       $ -0.071\pm0.021 $ & $ -0.80\pm0.09 $ & $ -0.097\pm0.025 $ &                   \\
 545 & $ -0.014\pm0.008 $ & $ +0.04\pm0.12 $ & $ -0.248\pm0.045 $ & $  0.000 $        &
       $  0.000         $ & $ -0.91\pm0.08 $ & $ -0.035\pm0.015 $ & $  0.000 $        \\
 602 & $ -0.113\pm0.015 $ & $ +0.21\pm0.14 $ & $ -0.114\pm0.052 $ & $  0.000 $        &
       $ -0.095\pm0.016 $ & $ -0.68\pm0.16 $ & $ -0.147\pm0.033 $ & $ -0.129\pm0.048$ \\
 710 & $ -0.119\pm0.049 $ & $ +0.23\pm0.08 $ & $ -0.119\pm0.049 $ &                   &
                          &                  &                    &                   \\
 712 & $  0.000         $ & $ -0.40\pm0.06 $ & $  0.000         $ &                   &
                          &                  &                    &                   \\
 799 & $ -0.086\pm0.010 $ & $ +0.54\pm0.12 $ & $ -0.116\pm0.021 $ & $ -0.062\pm0.035$ &
       $ -0.039\pm0.008 $ & $ -0.64\pm0.09 $ & $ -0.036\pm0.019 $ & $ -0.098\pm0.031$ \\
 898 & $ -0.184\pm0.010 $ & $ +0.55\pm0.02 $ & $ -0.184\pm0.010 $ &                   &
       $  0.000         $ & $ -0.58\pm0.14 $ & $  0.000         $ &                   \\
1076 & $ -0.131\pm0.010 $ & $ +0.19\pm0.02 $ & $ -0.131\pm0.010 $
&                   &
       $ -0.038\pm0.020 $ & $ -0.85\pm0.05 $ & $ -0.038\pm0.020 $ &                   \\
1088 & $ -0.061\pm0.009 $ & $ +0.47\pm0.09 $ & $  0.000         $
&                   &
       $ -0.027\pm0.006 $ & $ -0.65\pm0.06 $ & $  0.000         $ &                   \\
\noalign{\smallskip} \hline \hline
\end{tabular}
\end{flushleft}
}
 \label{TGRAD}
\end{table*}

\begin{figure}
\resizebox{\hsize}{!}{\includegraphics{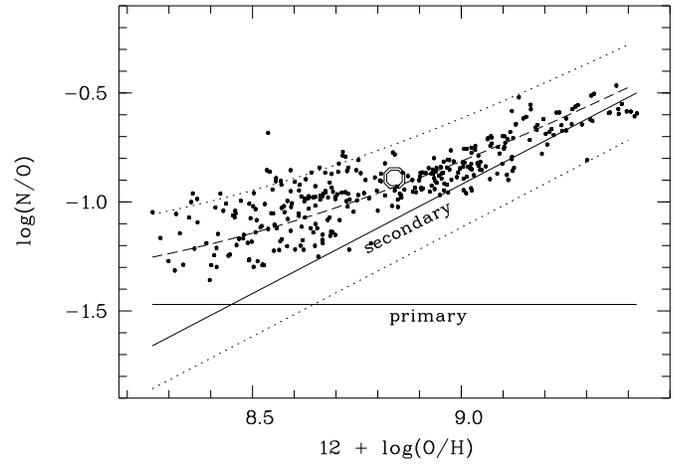}} \caption{N/O
abundance ratio as a function of oxygen abundance for the \hii\
regions in the sample galaxies. The different lines indicate the
relation for $primary$, $secondary$ and $primary + secondary$\
(dashed curve) origin of nitrogen (Vila--Costas \& Edmunds 1993).
The solar value is indicated by a large double circle. The two
dotted curves are envelopes corresponding to the observational
errors.}
  \label{no_vs_oh}
\end{figure}

With the data at hand, it is possible to derive the nitrogen--to--oxygen
abundance ratio (N/O).
For \hii\ regions in our sample, the N/O abundance ratio has been
derived from the dereddened emission lines \oii, \oiiib\ and \niib\ following
the method of Thurston \etal \cite*{THURSTONetal96}, as described in Coziol
\etal\ \cite*{COZIOLetal99a}.

The N/O abundance ratio as a function of oxygen abundance is shown in
Fig.~\ref{no_vs_oh}. Also shown in this figure are the expected relations for
a $primary$, a $secondary$ and a $primary + secondary$ origin for the
production of nitrogen \cite{V-CE93}. The \hii\ regions in our sample seem to
follow the $primary + secondary$ relation. But the evolution of N/O with
oxygen abundance does not trace a continuous behavior: it suddenly rises and
forms a sort of plateau between $8.7 < 12 + \mbox{log(O/H)} < 9.1$. This
behavior has been interpreted as evidence for chemical evolution by a sequence
of bursts of star formation (Coziol \etal\ 1999a).

\section{Results}

\subsection{Spatial distribution of oxygen abundances and of H$\alpha$}
\label{spatdist}

\begin{figure*}
\centering{\resizebox{16cm}{!}{\includegraphics{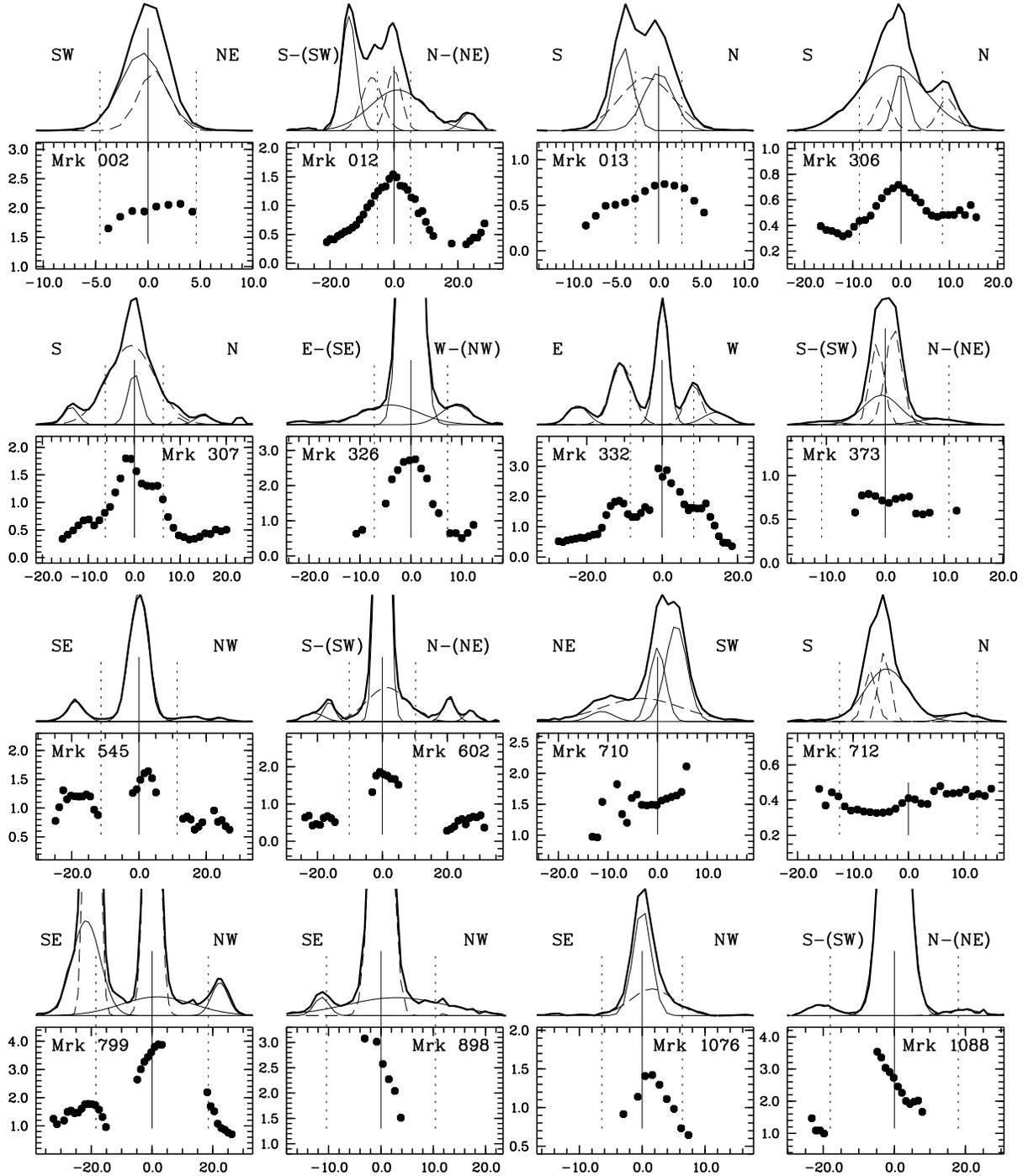}}}
  \caption{The oxygen abundance and \halpha\ emission of the galaxies
versus position along the slit in arcsec. The oxygen abundance is given in
solar units, adopting log(O/H) $\ = -3.16$\ for the solar value (Grevesse \&
Noels 1993). For each galaxy, a vertical continuous line marks its geometrical
center and two vertical dotted lines indicate the two ends of the bar.
In Mrk 710 the ends of the bar fall outside the figure (at 33.7 arcsec).
When the spatial \halpha\ profile is not a pure gaussian,  we also draw its
decomposition into multi--gaussian components}
  \label{oh_24gal}
\end{figure*}

\begin{figure*}
\centering{\resizebox{15cm}{!}{\includegraphics{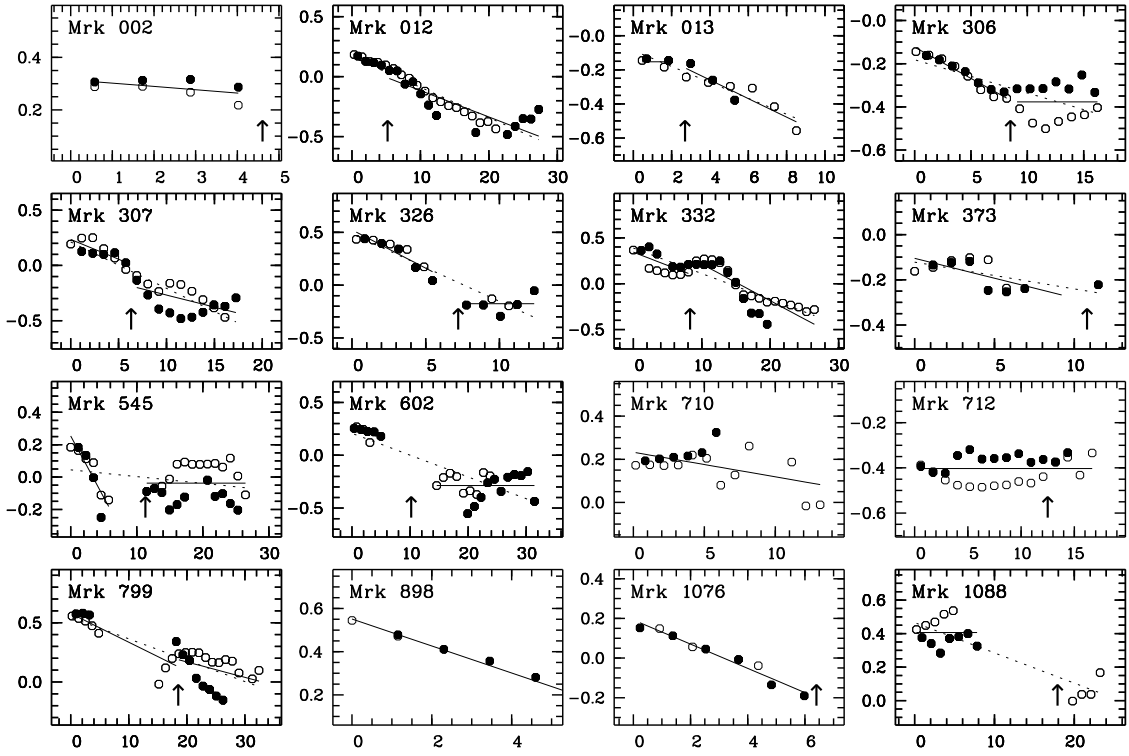}}}
  \caption{
Radial profiles of oxygen abundance log(O/H) in solar units. The radial
coordinate is in arcsec, with the
origin at the center of symmetry of the profile.
Values on both sides of the center are distinguished by two different symbols.
In each panel the vertical arrow indicates the end of the bar. The slopes of
the lines adjusted by least--square fits are given in Table~\ref{TGRAD}. The
two continuous lines correspond to the bar and disk abundance gradients and
the dotted line corresponds to the global gradient}
  \label{pli_oh_20gal}
\end{figure*}
\begin{figure*}
\centering{\resizebox{15cm}{!}{\includegraphics{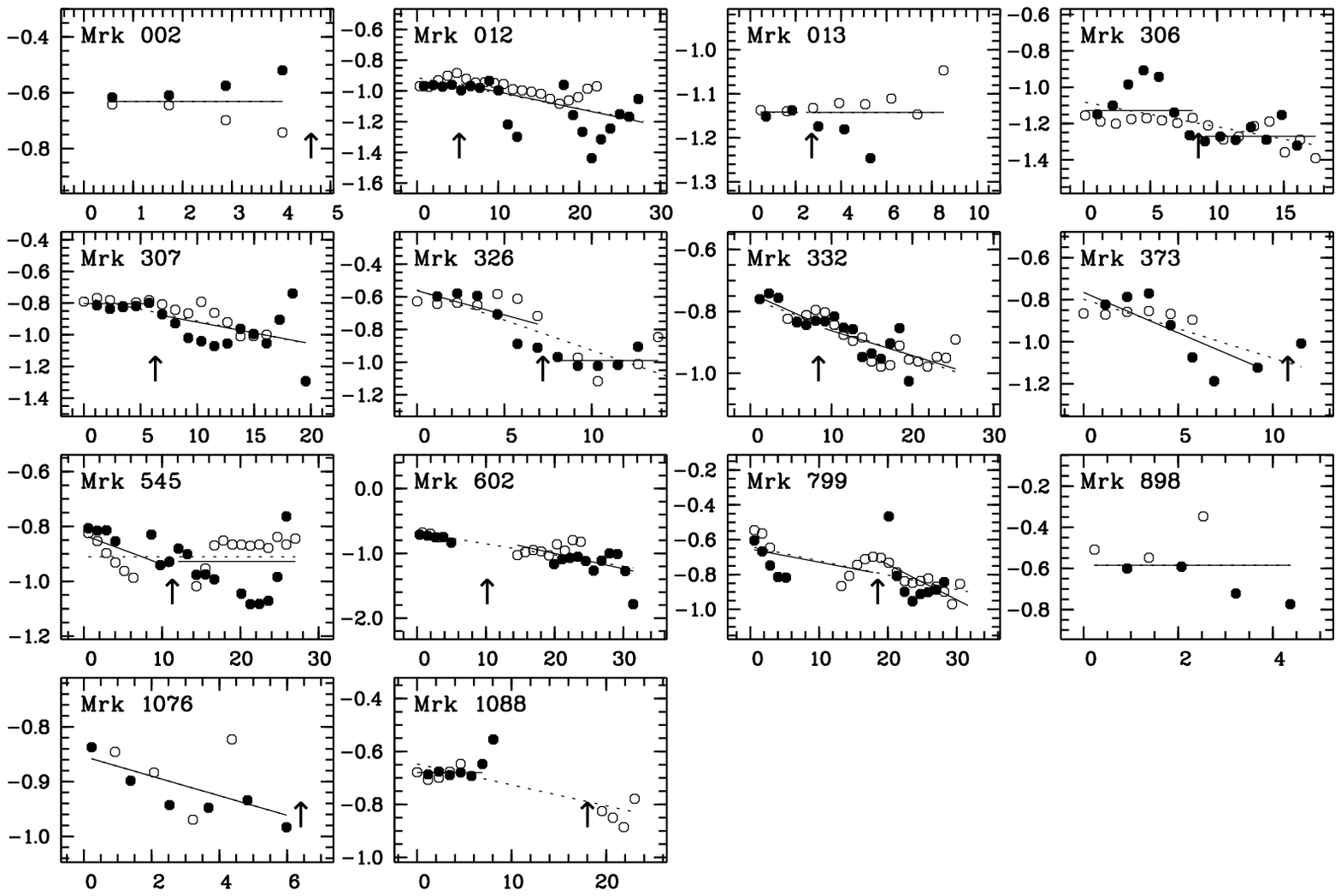}} }
  \caption{
Radial profiles of nitrogen-to-oxygen abundance ratio log(N/O).
The solar value on this scale
corresponds to $-0.9$. The meaning of the symbols is the same as in
Fig.~\ref{pli_oh_20gal}. The slopes of the lines adjusted by least--square fits
are given in Table~\ref{TGRAD}
}
  \label{pli_no_20gal}
\end{figure*}

The oxygen abundance and \halpha\ emission as a function of position along the
slit are presented in Fig.~\ref{oh_24gal}. The emission peak is usually
centered on the nucleus of galaxies. Exceptions are Mrk 12, 13, 712 and
799. In the case of Mrk 712 and 799, the maximum intensities correspond to
very young star--forming regions containing numerous Wolf--Rayet stars.  When
studying the gas distribution inside the bar, we find two possible situations:
more than half of galaxies have ionized gas covering the whole bar,
while in the rest of galaxies the ionized gas is mostly confined to the
center. A large number of galaxies (62\%) also show ionized gas either at
the two ends of the bar (25\%), or, more frequently, only at one end (37\%).

Examining the spatial distribution of the oxygen abundance, we find that it
usually reaches a maximum in the center of galaxies. In general, this
maximum corresponds to that of the \halpha\ emission.
In some galaxies, there is also strong
nebular emission outside the nucleus, accompanied by a local rise in oxygen
abundance (e.g. Mrk 306, 332 and 545). In other galaxies this secondary peak
of emission is not accompanied by an increase in oxygen abundance (e.g. Mrk
12, 13 and 799), and in some galaxies the oxygen abundance stays high even
without any secondary peak of \halpha\ emission (e.g. Mrk 307, 373 and 712).

\subsection{O/H and N/O abundance gradients}
\label{o_n_grad}

To study the abundance gradients, we have folded the profiles of O/H and
N/O abundance ratios about a center of symmetry. This center is the
locus of the principal peak of \halpha\ emission (see Fig.~\ref{oh_24gal}).

To make the analysis easier, we adjusted lines by least--square fits to the
data in two different regions of each profile: one includes the bar and the
nucleus and another the region outside the bar. The latter region corresponds
to the  disk. Note, however, that the profiles never cover the entire length
of the disk (that is out to a surface brightness of 25 \magarc), simply
because we did not detect ionized gas that far. A global gradient is also
estimated by fitting a line to the data over the full extent of the ionized
gas distribution. The intersect (central) abundance is determined by
extrapolating this global gradient up to the center. The two regressions are
plotted as continuous lines and the global gradient as a dotted line in
Figs.~\ref{pli_oh_20gal} and \ref{pli_no_20gal}. The profiles on the two sides
of the center are distinguished by different symbols. The numerical values of
different gradients and the intersects are given in Table~\ref{TGRAD}. The
uncertainties reported in that Table are those obtained from the regression
calculations.

The oxygen abundance profiles as a function of radius (in arcsec) are shown in
Fig.~\ref{pli_oh_20gal}. They show complicated patterns. Although the
uncertainties on the oxygen abundance ($\sim 0.2$ dex) prevent us from
interpreting the multiple inflections of these profiles, we believe that part
of these features must be real, reflecting variations in oxygen abundance
produced by different intensities of star formation along the bar. One good
example is Mrk 332, where there is an increase of oxygen abundance linked to
star formation at the two ends of the bar.

The abundance gradients (continuous and dotted lines in
Fig.~\ref{pli_oh_20gal}) also show various patterns. In Mrk 2, 373, 710, 712,
898 and 1076, the gradient can be estimated in the bar only. In these cases,
the bar gradient is considered the global gradient. Half of these gradients
are relatively shallow ($\sim -0.03$) and half are moderately strong
($\sim -0.15$).

For galaxies with ionized gas also in the disk, we distinguish two cases: the
disk abundance gradient can be either flat or negative. In Mrk 306, 326, 545
and 602, it is flat. Considering that normal spiral galaxies usually have
monotonous negative gradients \cite{ZARITSKYetal94}, this makes these galaxies
with flat disk gradient somewhat peculiar, although shallower outer profiles
have been observed before, for instances in NGC 1365 \cite{RW98}, NGC 3319
\cite{ZARITSKYetal94} and NGC 3359 \cite{MR95}. Other galaxies with
ionized gas in the disk look more ``normal'', with negative disk gradients. In
general, the gradients are shallower in the disk than in the bar. There are
two exceptions, Mrk 13 and Mrk 332. In the case of Mrk 332, the intense star
formation located at the ends of the bar is obviously responsible for the
steeper gradient in the disk (see Fig.~\ref{pli_oh_20gal}). For Mrk 13, there
is no gradient in the bar simply because this bar is very short, it is the
shortest bar among the 125 ones measured by Chapelon et al.
\cite*{CHAPELONetal99} (paper V). The bar of Mrk 1088 also seems to have a
flat gradient (Fig.~\ref{pli_oh_20gal}), but the unfolded oxygen abundance
profile (Fig.~\ref{oh_24gal}) shows that there is in fact a steep gradient
{\it accross} the center.

In general, our global abundance gradients are negative. This reflects
the fact that the oxygen abundance is almost always highest in the
center of the galaxy. One exception is Mrk 712, which does not have
any gradient at all.

We determined the N/O abundance ratios of all galaxies except Mrk 710 and
712 (see Sect.~\ref{nitroabund}).  The N/O profiles
are shown in Fig.~\ref{pli_no_20gal}. In general, they look
similar to those of oxygen.  The N/O abundance ratio also increases
toward the nucleus of the starburst galaxies, but the gradients are
significantly shallower and many more galaxies have zero gradient.
Further comparisons of
the two abundance gradients are postponed to Sect.~\ref{compabund}. Values for
the different gradients and the intersects are given in Table~\ref{TGRAD}.

\section{Analysis of the abundance gradients}
\label{compabund}

\subsection{Oxygen abundance gradients}

The presence of a bar could affect the abundance gradients in our sample
of galaxies in
two ways. By funneling metal--poor gas from the outer regions toward the
center, the bar is expected to dilute the chemical elements, and thus reduce
any initially present radial abundance gradient. But a bar could also
stimulate star formation at the centre, which, if strong enough,
could establish or
maintain a steep abundance gradient.  In the following analysis, we look for
such possible effects of the bar in our sample of starburst galaxies.

\begin{figure}
\hskip 0.8cm
\resizebox{!}{14cm}{\includegraphics{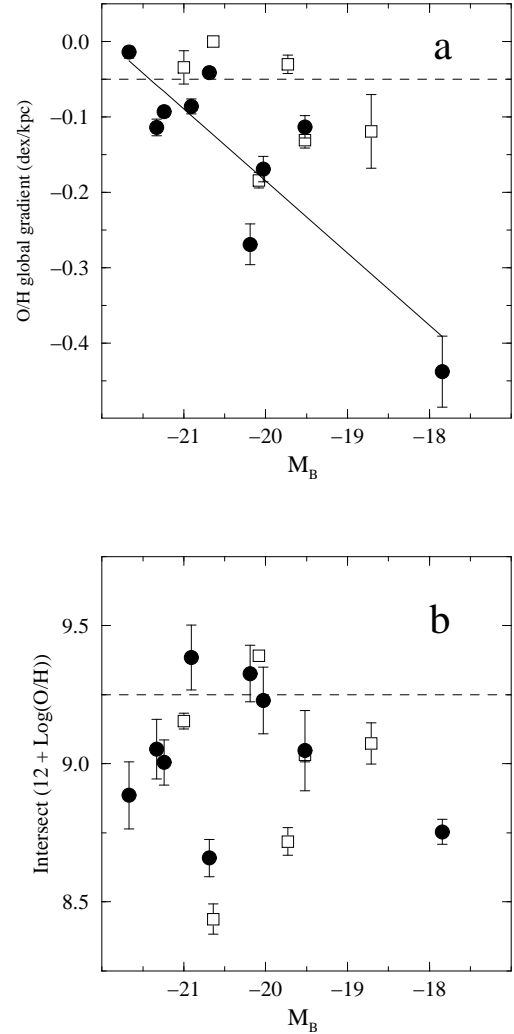}} \caption{ {\bf a}
Global oxygen abundance gradient as a function of absolute
magnitude. Galaxies where the global gradient is also the bar
gradient are indicated by open squares; filled circles represent
the other cases. The dashed line marks the mean gradient of normal
unbarred galaxies with absolute B magnitude between --19 and --22
in the sample of Edmunds \& Roy (1993). The continuous curve is a
linear regression on the filled circles (correlation coefficient
$= 86$\%). {\bf b} Intersect as a function of the absolute
magnitude. The meaning of the symbols is the same as in {\bf a}.
The dashed line indicates a mean value of the central oxygen
abundance in normal unbarred galaxies with absolute B magnitude
between --19 and --22 (from Edmunds \& Roy 1993) }
  \label{OH_Ggrad_Mb}
\end{figure}

Of the three categories of oxygen abundance gradients estimated here, the
global gradient is the one which is most comparable to those from other
studies. We show in Fig.~\ref{OH_Ggrad_Mb} the relation of global gradients
and intersects with absolute magnitudes. This Figure should be compared with
Fig.1 of Edmunds \& Roy (1993) for a sample of normal galaxies. The barred
starburst galaxies generally have steeper gradients than normal barred
galaxies \cite{EdR93,ZARITSKYetal94}. On average, the gradients in the barred
starburst galaxies are even steeper than those measured by
Edmunds \& Roy (1993) in
normal unbarred galaxies of comparable luminosity.  This shows that the
oxygen abundance in the sample of barred starburst galaxies is not diluted
by the mixing effect of the bar.

In the above analysis, six of the global abundance gradients are in fact bar
gradients, because no ionized gas was observed in the disks of these galaxies.
We should therefore reexamine Fig.~\ref{OH_Ggrad_Mb} and distinguish these
cases from those where the global gradient includes the disk gradient.  With
this distinction, we find that the global gradient in Fig.~\ref{OH_Ggrad_Mb}a
tends to steepen as the absolute magnitude increases.  A linear fit performed
on the data yields a correlation coefficient of 86\%.  But if we exclude the
extreme point (Mrk 13) at $M_{\rm B} \sim -18$, this coefficient drops to
56\%.  This is more a trend than a true correlation. Galaxies for which
the global gradient is the bar gradient also show a similar trend, but
slightly displaced towards lower values. An explanation for this trend is
proposed in Sect.~\ref{barsf}.

Examining now the central oxygen abundances (Fig.~\ref{OH_Ggrad_Mb}b), we find
that they are on average lower in barred starburst galaxies than in normal
unbarred galaxies. Because the global abundance gradients in barred starburst
galaxies are
strong, these low central abundances cannot be attributed to an artifact of
the bar, that would have lowered the central abundance of chemically evolved
galaxies by dilution. This means that the starburst galaxies in this sample have
lower oxygen abundances than normal ones.  This is in agreement with our
previous finding \cite{COZIOLetal97a,COZIOLetal98}:  SBNGs are chemically less
evolved than normal galaxies with comparable luminosity and morphology.  We
have shown elsewhere \cite{COZIOLetal97a,COZIOLetal98}, that the only way to
explain this phenomenon is by assuming that SBNGs are ``young'' galaxies.

\subsection{N/O abundance gradients}

We have shown that trends in the oxygen abundance gradients are
inconsistent with the mixing effect of a bar, unless the bar also triggers a
central starburst which more than compensates for the dilution.  However, the
fact that we find both strong oxygen abundance gradients and lower intersects
than in normal galaxies rules out a bar--triggered burst in the center of an
``old'' galaxy.

But the bar could be responsible for starbursts in the center of a relatively
``young'' galaxy. In this case, there are two alternatives. The gas funneled
by the bar could have been fed directly into the central region in a short
time (a few Myrs), prompting a new burst of star formation that is
sufficiently strong to produce steep abundance gradients. The other
alternative is that the bar acts over a much longer time scale
(a few Gyrs), progressively feeding the inner regions with gas that
is immediately recycled (since no mixing effects are seen).

We can test the above two alternatives by studying the N/O abundance
gradients and comparing them with the oxygen ones.  In Coziol \etal
\cite*{COZIOLetal99a}, we have shown that the bulk of nitrogen in SBNGs is
probably produced by intermediate--mass stars, whereas oxygen is mainly
released by massive stars.  In this scenario, the time scale for enrichment in
nitrogen ($400\ \mbox{Myr}\ < \tau < 2\ \mbox{Gyr}$) is
much longer than the one for oxygen
($\tau < 20$\ Myr). This allows us to look for effects of the bar on a time
scale covering a few Gyrs.

\begin{figure}
\hskip 0.8cm
\resizebox{!}{14cm}{\includegraphics{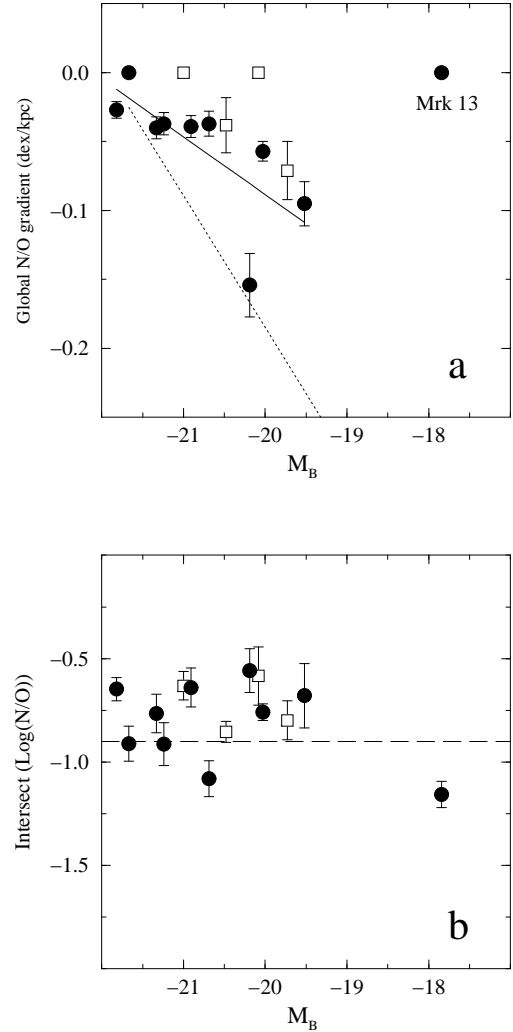}} \caption{ {\bf a}
Global N/O gradient as a function of absolute magnitude. The
symbols have the same meaning as in Fig.~\ref{OH_Ggrad_Mb}. The
continuous line is a regression on the N/O values, with a
correlation coefficient of 75\%. For the sake of comparison, we
also give the linear regression (dotted line) obtained for the
gradient in O/H (see Fig.~\ref{OH_Ggrad_Mb}). {\bf b} Intersect as
a function of absolute magnitude. The horizontal dashed line
corresponds to the solar value of N/O}
  \label{NO_Ggrad_Mb}
\end{figure}

In Fig~\ref{NO_Ggrad_Mb}a, we show the relation between the global N/O
abundance gradient and absolute magnitude.  In this Figure, we distinguish
galaxies with a gradient measured only along the bar from those where the
gradient also includes the disk.  We observe the same trend for N/O as
for oxygen: galaxies with a lower luminosity seem to have a stronger
negative gradient.  The relation implied by this tendency has a weak
correlation coefficient (60\% with Mrk 13 excluded) and a shallower slope than
that found for oxygen.
Again, galaxies with only a bar gradient seem to have
shallower gradients than the other ones.

Despite the small gradients, the central N/O values, shown in
Fig.~\ref{NO_Ggrad_Mb}b, are rather high compared to the solar abundance.
This implies that many generations of intermediate--mass stars have already
contributed to enrich these galaxies in nitrogen, or, equivalently, that enough
time has passed in the chemical evolution of these galaxies.

Why is the gradient systematically weaker for N/O than for
oxygen?  If this is due to the bar, it means that the effect of the bar
appears on a time scale comparable to the nitrogen enrichment phase.
This is unlikely as it implies that the oxygen enrichment phase, and the
ensuing
strong oxygen abundance gradient, happened after the nitrogen one.  Therefore,
as a mixing mechanism, the bar should have affected both elements in the same
way, which is not observed.  Our explanation for the observed
difference is given in Sect~\ref{constgrad}.

\begin{figure}
\hskip 0.8cm
\resizebox{!}{14cm}{\includegraphics{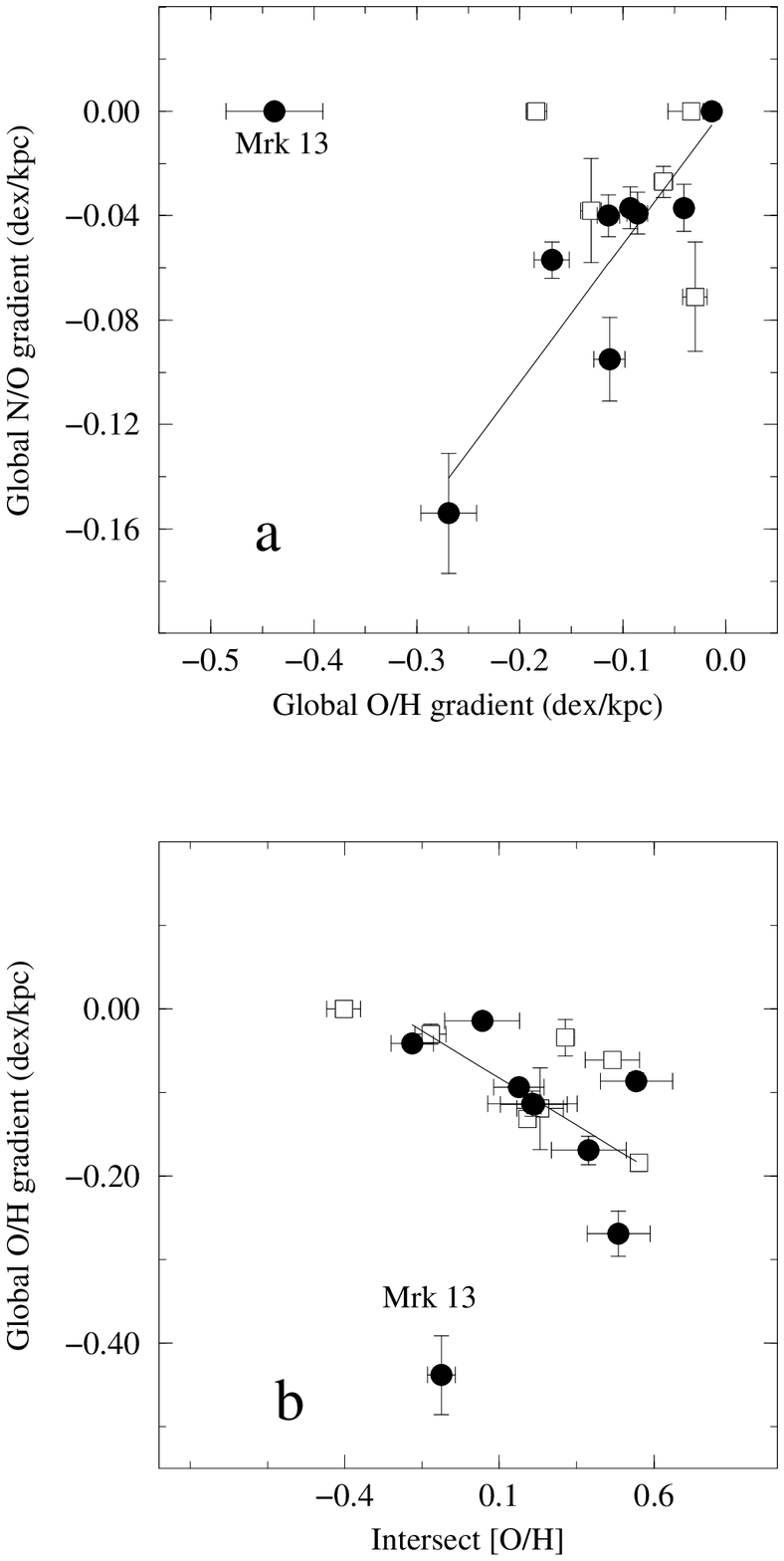}} \caption{ {\bf a}
Relation between global gradients in N/O  and O/H. The symbols
have the same meaning as in Fig.~\ref{OH_Ggrad_Mb}. The continuous
curve is a linear regression with a correlation coefficient of
89\%. {\bf b} Relation between the global gradients in O/H and its
intersect. The continuous curve is a linear regression with a
correlation coefficient of 68\% }
  \label{NO_vs_OH_Ggrad}
\end{figure}

The fact that the N/O intersects are high rules out the possibility that the
bars recently triggered central starbursts in these galaxies. A recent burst
would have increased the oxygen abundance and decreased the N/O abundance
ratio. The stronger the burst, the stronger the decrease in N/O.
We would thus expect galaxies with the steepest oxygen abundance gradients
to have the flattest N/O ones. In Fig.~\ref{NO_vs_OH_Ggrad}a, we show
the relation between the oxygen and N/O abundance gradients.
The behavior is contrary to what is expected.
The bars consequently did not recently trigger bursts in these galaxies.

\subsection{The progressive build up of abundance gradients}
\label {constgrad}

What is then the reason for the difference between the two gradients?
In fact, the N/O abundance gradient is not independent of the oxygen
one.  This is shown in Fig.~\ref{NO_vs_OH_Ggrad}a.  A linear regression on the
data yields the relation:  $\Delta({\rm N/O})/\Delta {\rm r}\ =
0.53\ (\pm0.10)\ \Delta({\rm O/H})/\Delta {\rm r}\ + 0.00\ (\pm0.01)$\
with a correlation coefficient of 89\%.
Surprisingly, the slope of this relation is almost the same as the one found
between the N/O and oxygen abundances:  a linear fit performed on the
data in Fig.~\ref{no_vs_oh} yields log(N/O)$\ =
0.55$log(O/H)$\ +0.8$ \cite{COZIOLetal99a}, which is consistent with a mixture
of $primary + secondary$ origin for nitrogen \cite{McGaugh91}.

The relation between the two gradients is thus explained by the
behavior of the oxygen abundance in the nuclei of the galaxies.  This is shown
in Fig.~\ref{NO_vs_OH_Ggrad}b, where the oxygen abundance gradient steepens
when its intersect increases. This relation is trivial if we assume that all
disks initially have about the same low oxygen abundance (see
Fig.~\ref{pli_oh_20gal}).  Since the nitrogen enrichment is related to the
increase in oxygen (See Sect.~\ref{nitroabund} and Fig.~\ref{no_vs_oh}), the
N/O abundance gradient grows with that of oxygen.  The origin of both
gradients in starburst galaxies can thus be explained solely by their star
formation histories.

\section{The role of bars in starburst galaxies}

The only way to reconcile the above observations of abundance gradients with
the possible effect of a bar is by assuming that the latter is a slow
process; over a few Gyr period, the bar progressively feeds the nucleus with gas which
is immediately recycled into stars.
This scenario requires a very fine tuning between the
feeding of the gas in the center of galaxies and star formation. It also
requires that bars are relatively stable over a long period of time (a few
Gyrs). We can test if these conditions apply to galaxies in our sample by
looking for relations between the bar properties and star formation.

\subsection{Bar properties and star formation: evidence for young bars}

\begin{figure}
\hskip 0.8cm
{\centering\resizebox{!}{10cm}{\includegraphics{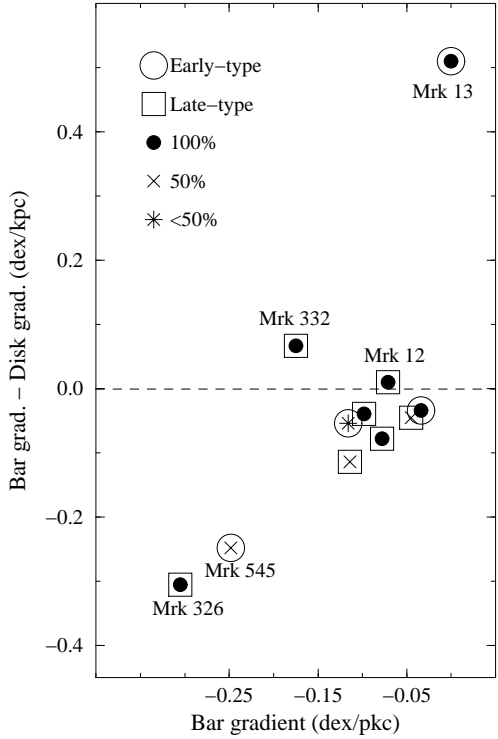}}}
\caption{Difference between oxygen abundance gradients in the bar
and in the disk. The difference is shown as a function of the
gradient in the bar, which explains why the galaxies with a bar
gradient only follow a 1 to 1 relation. The galaxies with ionized
gas over the whole bar ($\bullet$) are distinguished from those
where the gas covers 50\% ($\times$) or less ($\ast$) of the bar.
Early--type galaxies are indicated by a large open circle and
late--type ones by a large open square}
  \label{BvsDgrad}
\end{figure}

\begin{figure}
\hskip 0.6cm
{\centering\resizebox{!}{6.5cm}{\includegraphics{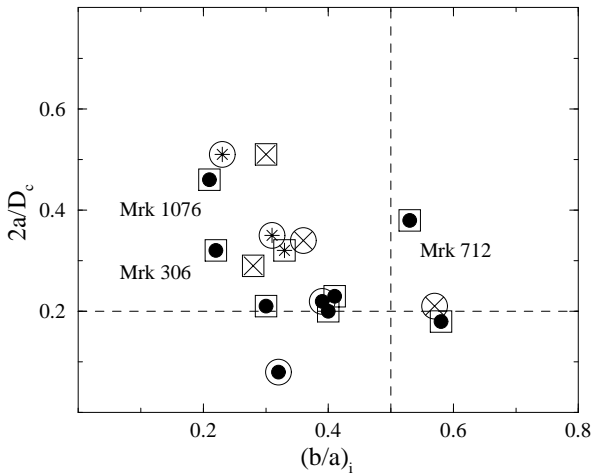}}}
\caption{ Distribution of ionized gas along the bar as a function
of bar and galaxy morphologies. The symbols have the same meaning
as in Fig.~\ref{BvsDgrad}. The horizontal and vertical dashed
lines indicate the limits of short and weak bars respectively.
There is a tendency for the ionized gas to be less extended in
longer bars }
  \label{gas_distribution}
\end{figure}

Numerical simulations predict that a strong bar maintains a steeper abundance
gradient in the bar than in the disk for about 1 Gyr, then both become
comparably low ($\sim -0.02$), and that the bar and disk gradients are never
very different in the presence of a weak bar \cite{MF97}.
We show in Fig.~\ref{BvsDgrad} that almost all galaxies in our sample show a
stronger gradient in the
bar than in the disk.  Only two galaxies, Mrk 13 and 332, have significantly
steeper gradients in the disk than in the bar, but this peculiarity can be
explained by other phenomena (see Sect.~\ref{o_n_grad}). In other words, the
difference between bar and disk gradients cannot be due to an old bar.
It cannot be due to the effect of a young bar either, because,
as we previously noted, the N/O abundance ratios would be low
and the gradients would be anti--correlated.

A simple explanation for the stronger gradients in the bar than in the disk is
that the volume density of star formation is proportional to the gas
volume density to some power $n$ ($1\le n \le 2$), as proposed by
Schmidt's law (1959, 1963).  Assuming that the chemical enrichment
is proportional to the star formation, a spherical density distribution
yields a variation with radius in
$R^{-3/n}$.  The gradient measured from the center to the end of the bar
(or to any intermediate radius) will
necessarily be steeper than the one measured further out.

The presence of ionized gas is a tracer of star formation in galaxies.
Half the galaxies in this sample show the presence of ionized gas over the
whole bar, while in the
other half, the gas is mostly concentrated in the inner parts of the bar
and in the nucleus.  If star formation cannot result from gas flow
toward the
inner regions by the dynamical action of a bar, could it be linked to any
other property of the bar, such as its strength or its length? The bar
strength is identified with the deprojected bar axis ratio (b/a)$_i$, and its
length is given by the ratio of the deprojected major axis to corrected blue
isophotal diameter 2a/D$_c$ \cite{CHAPELONetal99} (Paper V). A bar with a
ratio b/a $< 0.5$ is considered ``strong'' and a bar with a ratio 2a/D$_c >
0.2$ is considered ``long''.  We also identify separately galaxies with
various fractions of ionized gas along the bar and galaxies with early--
(earlier than SBbc) or late--type (SBbc or later) morphologies.

As shown in Fig.~\ref{gas_distribution}, most galaxies in this sample have a
strong bar, which is a general characteristic of the sample of Markarian
barred starburst galaxies \cite{CHAPELONetal99}.  There is a trend for
galaxies with a longer bar to have ionized gas over a smaller extent of the
bar length, regardless of their morphology. This is not an observational
selection effect, because three galaxies do not follow the above trend in
Fig.~\ref{gas_distribution}, namely Mrk 306, 712 and 1076. In the case of Mrk
306 and 1076, a long and strong bar with ionized gas over its whole length may
be explained by the fact that they are both interacting with close companions
(the rest of our galaxies are isolated). Mrk 712 looks like an isolated
galaxy, but with a peculiar morphology. The peak of star forming
activity is significantly displaced from the center of the galaxy (see
Fig.~\ref{oh_24gal}). This active region may be associated with a second
nucleus \cite{MB93}, which makes this galaxy a strong candidate for a recent
merger. A simple relation with the length of the bar thus explains the radial
distribution of ionized gas; star formation does not cover the whole bar
just because it is too long.

We have shown so far that bars, despite the fact that they are
strong, cannot account for the observed abundance gradients.  This may
be surprising in view of the convincing evidence for dynamical effects
of strong bars on star formation and abundance gradients in galaxies
provided by numerical simulations.  The only explanation for such a
situation is that bars in the current sample of galaxies
are too young (a few $10^7$ years) to have had any effect.

We verified that there are no relations between the
abundance gradients and the bar properties.
The oxygen and N/O abundance gradients in the bar and in the
disk show the same behavior. No relation is found between the abundance
gradients and the bar strength or length. Nor is any relation found between
the abundance gradients and the different concentrations of gas along the bar.
In general, therefore, the bars have not influenced the two gradients.
This supports our assumption that the bars are young.

\subsection{Bar--induced star formation at the bar ends}
\label{barsf}

If we compare the distribution of ionized gas (i.e. star formation) with the
oxygen abundance, we find that the behaviors of these two parameters along the
bars are frequently uncorrelated (see Fig.~\ref{oh_24gal} and
Sect.~\ref{spatdist}). Intense star formation does not
necessarily enhance the oxygen abundance, as we can see,
for example, in the extranuclear star forming regions of Mrk 12 and 799.
These regions must be too young (a few Myrs)
and their massive stars did not had
enough time to change the chemical abundance of their environment.
When we find, on the other hand, a region
filled with ionized gas, but where the abundance is significantly higher, we
conclude that this must be a region where star formation was ``stable'' over a
longer period of time.  The galaxy nuclei in our sample are obvious
locations for such stable star-forming regions.  The relatively high N/O
abundance ratios and the high proportion of intermediate--mass stars mixed with
ionized gas indicate that star formation persisted in these regions for quite
a long time (a few Gyrs).

We found no evidence that bars triggered the starbursts observed at the
center of the sample galaxies.
On the other hand, they may have induced star formation at the ends of the bar
(e.g.  Mrk 307, 332), and more frequently at only one end (e.g. Mrk 12, 13,
306, 545 and 799). This phenomenon may be due to gas compression in these
regions \cite{RHA79}.  Star formation
induced by the bar at its ends has increased the oxygen
abundance there.  The result is a shallower
abundance gradient in the bar and a stronger one in the disk. This may
explain why the global oxygen and N/O abundance gradients are weaker when
they are measured only in the bar. Chemical enrichment
produced by star formation at these points being lower than what
is observed in the nucleus, this suggests that fewer star formation episodes
occurred there, or that they happened relatively recently.

Among all the possible effects of the bar on the evolution of starburst
galaxies, therefore, the only plausible one seems to be star formation induced
at one or both ends of the bars. The distribution of star formation in this
sample of starburst galaxies is consistent with the assumption that
bars are too young (a few $10^7$ years old) to be at the origin of the bursts
in the galaxy nuclei.

\section{Summary and conclusion}

One important result of our analysis is that bars
have not had a great impact on the star
formation history and chemical evolution of starburst galaxies.
The most straightforward explanation is that bars are too young.
Our observations also clearly show that these young bars cannot be at
the origin of the nuclear starbursts. The high N/O abundance ratios
together with the high level of star formation in the nuclear region of
galaxies suggest that star formation proceeds almost
steadily (or as a sequence of bursts)
over a few Gyr period.

The fact that we find strong O/H gradients while the oxygen
abundance is low in the galactic nuclei supports our interpretation
that SBNGs are young galaxies still in their
process of formation (Coziol et al. 1997a; 1997b).
It seems that these starburst galaxies are still building their
abundance gradients.
This process can be fully explained
in terms of their star formation history. The
gradients build up from the inside out, becoming stronger as the oxygen and
N/O abundances increase in the bulge while staying low in the disk.
This behavior is consistent with a simple Schmidt law relating
the density of star formation to that of gas.

According to these results, starburst galaxies seem to form their
bulge first, and then their disk.  If the disk is younger than the bulge and
the bar forms in the disk, it is not surprizing that we find mostly young bars
in starburst galaxies.
The above scenario may also explain why our results
are not consistent with predictions made by
different models of bar formation.
The initial conditions assumed in these models
are far from those observed in young starburst galaxies.
A bar forms in a galaxy which already has a large bulge but a disk
which is probably still in formation.

When star formation stops in the bulge and increases in the disk,
as in normal spiral galaxies, the metallicity
of the disk grows, decreasing the gradient.
If the end product of a starburst
galaxy is a normal spiral galaxy, the latter, therefore,
should have lower metallicity gradients than the former.
If they are stable, bars may also play a more
important role as galaxies get older.

But how general is our conclusion? Is what we observe a trait of SBNGs?
What is the relation between the bulge and disk formation in starburst
galaxies and in normal spiral galaxies? These are the questions
we address in our companion paper (Coziol et al. 2000).

\begin{acknowledgements}
We thank the staff of Observatoire de Haute--Provence for
assistance at the telescope. R. C. would also like to thank
Observatoire de Besan\c{c}on for funding his visit, during which
this paper was completed. He would also like to thank the
direction and staff of Observatoire de Besan\c{c}on for their
hospitality. We acknowledge with thanks from the referee, Danielle
Alloin, positive comments and constructive suggestions which have
helped to improve the quality of this paper. For this research, we
have made use of the Lyon-Meudon Extragalactic Database (LEDA),
operated by the Lyon and Paris-Meudon Observatories (France). We
also used the NASA/IPAC Extragalactic Database (NED) which is
operated by the Jet Propulsion Laboratory, California Institute of
Technology, under contract with the National Aeronautics and Space
Administration.

\end{acknowledgements}



\begin{thebibliography}{}

\bibitem[\protect\astroncite{Alloin \etal}{1979}]{ALLOINetal79}
Alloin D., Collin--Souffrin S., Joly M., Vigroux L., 1979, A\&A 78, 5

\bibitem[\protect\astroncite{Baldwin \etal}{1981}]{BPT81}
Baldwin J.A., Phillips M.M., Terlevich R., 1981, PASP 93, 5

\bibitem[\protect\astroncite{Chapelon \etal}{1999}]{CHAPELONetal99}
Chapelon S., Contini T., Davoust E., 1999, A\&A 345, 81 (Paper V)

\bibitem[\protect\astroncite{Contini}{1996}]{C96}
Contini T., 1996, Ph. D. thesis, Universit\'e Paul Sabatier,
Toulouse, France

\bibitem[\protect\astroncite{Contini \etal}{1995}]{CONTINIetal95}
Contini T., Davoust E., Consid\`ere S., 1995, A\&A 303, 440 (Paper I)

\bibitem[\protect\astroncite{Contini \etal}{1996}]{CONTINIetal96}
Contini T., Wozniak H., Consid\`ere S., Davoust E., 1996,
in Proceedings of the workshop ``The physics of LINERs in view
of recent observations", {\it ASP Conference series}, 103, 175

\bibitem[\protect\astroncite{Contini \etal}{1997}]{CONTINIetal97}
Contini T., Wozniak H., Consid\`ere S., Davoust E., 1997, A\&A 324, 41
(Paper II)

\bibitem[\protect\astroncite{Contini \etal}{1998}]{CONTINIetal98}
Contini T., Consid\`ere S., Davoust E., 1998, A\&AS 130, 285 (Paper III)

\bibitem[\protect\astroncite{Coziol}{1996}]{COZIOL96}
Coziol R., 1996, A\&A 309, 345

\bibitem[\protect\astroncite{Coziol \etal}{1997a}]{COZIOLetal97a}
Coziol R., Contini T., Davoust E., Consid\`ere S., 1997a, ApJ 481, L67

\bibitem[\protect\astroncite{Coziol \etal}{1997b}]{COZIOLetal97b}
Coziol R., Demers S., Barn\'eoud R., Pena M., 1997b, AJ 113, 1548

\bibitem[\protect\astroncite{Coziol \etal}{1998}]{COZIOLetal98}
Coziol R., Contini T., Davoust E., Consid\`ere S., 1998, in
Abundance Profiles: Diagnostic Tools for Galaxy History,
Eds.: D. Friedli, M. Edmunds, C. Robert, L. Drissen, ASP Conf. Ser.
Vol. 147, (San Francisco) p. 219.

\bibitem[\protect\astroncite{Coziol \etal}{1999a}]{COZIOLetal99a}
Coziol R., Carlos--Reyes R.E., Consid\`ere S., Davoust E., Contini T.,
1999a, A\&A 345, 733

\bibitem[\protect\astroncite{Coziol \etal}{1999b}]{COZIOLetal2000}
Coziol R., Consid\`ere S., Davoust E., Contini T.,
2000, submitted

\bibitem[\protect\astroncite{Edmunds}{1990}]{EDMUNDS90}
Edmunds M.G., 1990, MNRAS 246, 678

\bibitem[\protect\astroncite{Edmunds \& Pagel}{1984}]{ED84}
Edmunds M.G., Pagel B.E.P., 1984, MNRAS 211, 507

\bibitem[\protect\astroncite{Edmunds \& Roy}{1993}]{EdR93}
Edmunds M.G., Roy J.--R., 1993, MNRAS 261, L17

\bibitem[\protect\astroncite{Friedli \etal}{1994}]{FRIEDLIetal94}
Friedli D., Benz W., Kennicutt R., 1994, ApJ 430, L105

\bibitem[\protect\astroncite{Grevesse & Noels}{1993}]{GREVNO93}
Grevesse N., Noels A., 1993, in ``Origin and Evolution of the Elements'',
Prantzos N., Vangioni--Flam E., Cass\'e M. eds., Cambridge University Press,
p. 15

\bibitem[\protect\astroncite{G\"otz \& K\"oppen}{1992}]{GK92}
G\"otz M., K\"oppen J., 1992, A\&A 262, 455

\bibitem[\protect\astroncite{Heckman \etal}{1998}]{Hetal98}
Heckman T.M., Robert R., Leitherer C., Garnett D.R., van der Rydt F.,
1998, APJ 503, 646

\bibitem[\protect\astroncite{Lema\^\i tre \etal}{1990}]{LEMAITREetal90}
Lema\^\i tre G. Kohler D., Lacroix D., Meunier J.-P., Vin A., 1990,
A\&A 228, 540

\bibitem[\protect\astroncite{Martin \& Roy}{1994}]{MR94}
Martin P., Roy J.-R., 1994, ApJ 424, 599

\bibitem[\protect\astroncite{Martin \& Roy}{1995}]{MR95}
Martin P., Roy J.-R., 1995, ApJ 445, 161

\bibitem[\protect\astroncite{Martinet \& Friedli}{1997}]{MF97}
Martinet L., Friedli D., 1997, A\&A 323, 363

\bibitem[\protect\astroncite{Massey \etal}{1988}]{MASSEYetal88}
Massey P., Strobel K., Barnes J.V., Anderson E., 1988, ApJ 328, 315

\bibitem[\protect\astroncite{Mazzarella \& Boroson}{1993}]{MB93}
Mazzarella J.M., Boroson T.A., 1993, ApJS 85, 27

\bibitem[\protect\astroncite{McCall \etal}{1985}]{MCCALLetal85}
McCall M.L., Rybski P., Shields G.A., 1985, ApJS 57, 1

\bibitem[\protect\astroncite{McGaugh}{1991}]{McGaugh91}
McGaugh S., 1991, ApJ 380, 140

\bibitem[\protect\astroncite{Moll\'a \etal}{1996}]{MOLLAetal96}
Moll\'a M., Ferrini F., D\'\i az A.I.,  1996, ApJ 466, 668

\bibitem[\protect\astroncite{Noguchi}{1988}]{N88}
Noguchi M., 1988, A\&A 203, 259

\bibitem[\protect\astroncite{Oye \& Kennicutt}{1993}]{OK93}
Oye M.S., Kennicutt R.C.Jr., 1993, ApJ 411, 137

\bibitem[\protect\astroncite{Pagel}{1989}]{P89}
Pagel B., 1989, Rev. Mex. Astron. Astrof. 18, 161

\bibitem[\protect\astroncite{Roberts \etal}{1979}]{RHA79}
Roberts W.W., Huntley J.M, van Albada G.D., 1979, ApJ 233, 67

\bibitem[\protect\astroncite{Roy \& Walsh}{1998}]{RW98}
Roy J.-R., Walsh J.R., 1998, MNRAS 288, 715

\bibitem[\protect\astroncite{Schmidt}{1959}]{Schmidt59}
Schmidt M., 1959, ApJ 129, 243

\bibitem[\protect\astroncite{Schmidt}{1959}]{Schmidt63}
Schmidt M., 1963, ApJ 137, 758

\bibitem[\protect\astroncite{Thurston \etal}{1996}]{THURSTONetal96}
Thurston T.R., Edmunds M.G., Henry R.B.C., 1996, MNRAS 283, 990

\bibitem[\protect\astroncite{Vacca \& Conti}{1992}]{VC92}
Vacca W.D., Conti P.S., 1992, ApJ 401, 543

\bibitem[\protect\astroncite{Veilleux \& Osterbrock}{1987}]{VO87}
Veilleux S., Osterbrock D.E., 1987, ApJS 63, 295

\bibitem[\protect\astroncite{Vila--Costas \& Edmunds}{1992}]{V-CE92}
Vila--Costas M.B., Edmunds M.G., 1992, MNRAS 259, 121

\bibitem[\protect\astroncite{Vila--Costas \& Edmunds}{1993}]{V-CE93}
Vila--Costas M.B., Edmunds M.G., 1993, MNRAS 265, 199

\bibitem[\protect\astroncite{Wise \& Silk}{1989}]{WS89}
Wise R.F.G., Silk J., 1989, ApJ 339, 700

\bibitem[\protect\astroncite{Zaritsky \etal}{1994}]{ZARITSKYetal94}
Zaritsky D., Kennicutt R.C.Jr., Huchra J.P., 1994, ApJ 420, 87

\end{thebibliography}
\end{document}